\documentclass[aps,pra,twocolumn,superscriptaddress]{revtex4}
\usepackage{graphicx}
\usepackage{natbib}
\usepackage{amsmath}
\usepackage{amssymb}
\usepackage{epsfig}

\newcommand{\beq}{\begin{eqnarray}}
\newcommand{\eeq}{\end{eqnarray}}

\begin{document}

\title{Unified single-photon and single-electron counting statistics:\\
from cavity-QED to electron transport}%\footnote{If necessary, the
%more technical parts of this manuscript, and the Methods section,
%can be moved to an online supplementary section.}}
\author{Neill Lambert}
\affiliation{Advanced Science Institute, RIKEN, Saitama 351-0198,
Japan}

\author{Yueh-Nan Chen}
\affiliation{Department of Physics and National Center for
Theoretical Sciences, National Cheng-Kung University, Tainan 701,
Taiwan}

\author{Franco Nori}
\affiliation{Advanced Science Institute, RIKEN, Saitama 351-0198,
Japan}

\affiliation{Physics Department, The University of
Michigan, Ann Arbor, M1 48109-1040, USA}

\date{\today}
\maketitle

\textbf{A key ingredient of cavity quantum-electrodynamics (QED)
is the coupling between the discrete energy levels of an atom and
photons in a single-mode cavity. The addition of periodic
ultra-short laser pulses allows one to use such a system as a
source of single photons; a vital ingredient in quantum
information and optical computing schemes. Here, we analyze and
``time-adjust'' the photon-counting statistics of such a
single-photon source, and show that the photon statistics can be
described by a simple `transport-like' non-equilibrium model. We
then show that there is a one-to-one correspondence of this model
to that of non-equilibrium transport of electrons through a double
quantum dot nanostructure. Then we prove that the statistics of
the tunnelling electrons is equivalent to the statistics of the
emitted photons. This represents a unification of the fields of
photon counting statistics and electron transport statistics. This
correspondence empowers us to adapt several tools previously used
for detecting quantum behavior in electron transport systems
(e.g., super-Poissonian shot noise, and an extension of the
Leggett-Garg inequality) to single-photon-source experiments.}

Cavity QED studies the interaction between a two-level atom and a
single-mode cavity (see, e.g., refs.~1-4). Vacuum Rabi
oscillations, the coherent excitation transfer between atoms and
cavity photons, can occur if the atom-photon coupling strength $g$
overwhelms both the loss rate ($\kappa $) of the cavity photons
and the emission rate ($\gamma $) into other modes, as shown
schematically in Fig. 1. To observe such a quantum oscillation, a
velocity-selected atomic beam is passed through an open
Fabry-Perot
resonator to control the interaction time $t_{i}$. The probability $%
P_{e}(t_{i})$ that the atom remains in the excited state $\left|
e\right\rangle $ at time $t_{i}$ is written$^{1-3}$ as
$P_{e}=(1+\cos 2gt_{i})/2$. This coherent coupling can be observed
by examining the so-called vacuum Rabi splitting (VRS) in the
transmission spectrum of the cavity. Clear evidence of VRS has
been demonstrated not only in atomic
systems$^{2-4}$, but also in semiconductor self-assembled quantum dots$^{5}$%
(QD) and circuit QED$^{6-8}$ systems.

In several recent experiments (see, e.g., refs.~1-4), a cavity QED
system was used as a source of single photons by deterministically
exciting the atom via periodic ultra-short laser pulses. Normally,
one interprets the total photon statistics from such an experiment
as the ensemble average of a single event: the atom is excited at
$t=0$, interacts with the cavity, and eventually the cavity photon
is emitted at some later time. All of the recorded single-photon
detection events are then combined to give the ensemble average of
this single situation.

Here, we propose a simple alternative method of analyzing the photon
detection events of this kind of experiments$^{1-4}$ that we term
``time-adjusted photon counting''. We show that the photon emission spectrum
can then be modelled via a Markovian master equation which has a one-to-one
correspondence to a well studied model of a double quantum dot (DQD) in the
large bias, Coulomb-blockade regime. This allows us to reinterpret data from
existing (and future) cavity QED single-photon-source experiments as a
continuous ``transport-like'' phenomenon, unifying photon and electron
statistics. A summary of this correspondence can be found in Table I.

Double quantum dots are artificial atoms in a solid$^{9-11}$. A
variety of powerful tools have been developed to study transport
through such devices. These tools have revealed unique features
like Coulomb-blockade$^{12}$, the Kondo effect$^{12}$, and
coherent oscillations$^{13,14}$. Our main result here is that the
\emph{electron-counting} statistics developed for the DQD model
(e.g., current, current-noise, and higher-order cumulants) can be
observed in existing \emph{photon-counting} cavity QED
experiments. We will use the analogy between these two
apparently-unrelated systems to show that the photon statistics
has a non-negative shot-noise feature, complementary to their
sub-Poissonian anti-bunching statistics, that indicates VRS.
Moreover, we calculate the second-order correlation functions and
show that these violate an extended form$^{15}$ of the
Leggett-Garg inequality$^{16}$. For completeness, we also consider
violations of this inequality by the non-adjusted statistics.

\begin{figure}[h]
\includegraphics[width=\columnwidth]{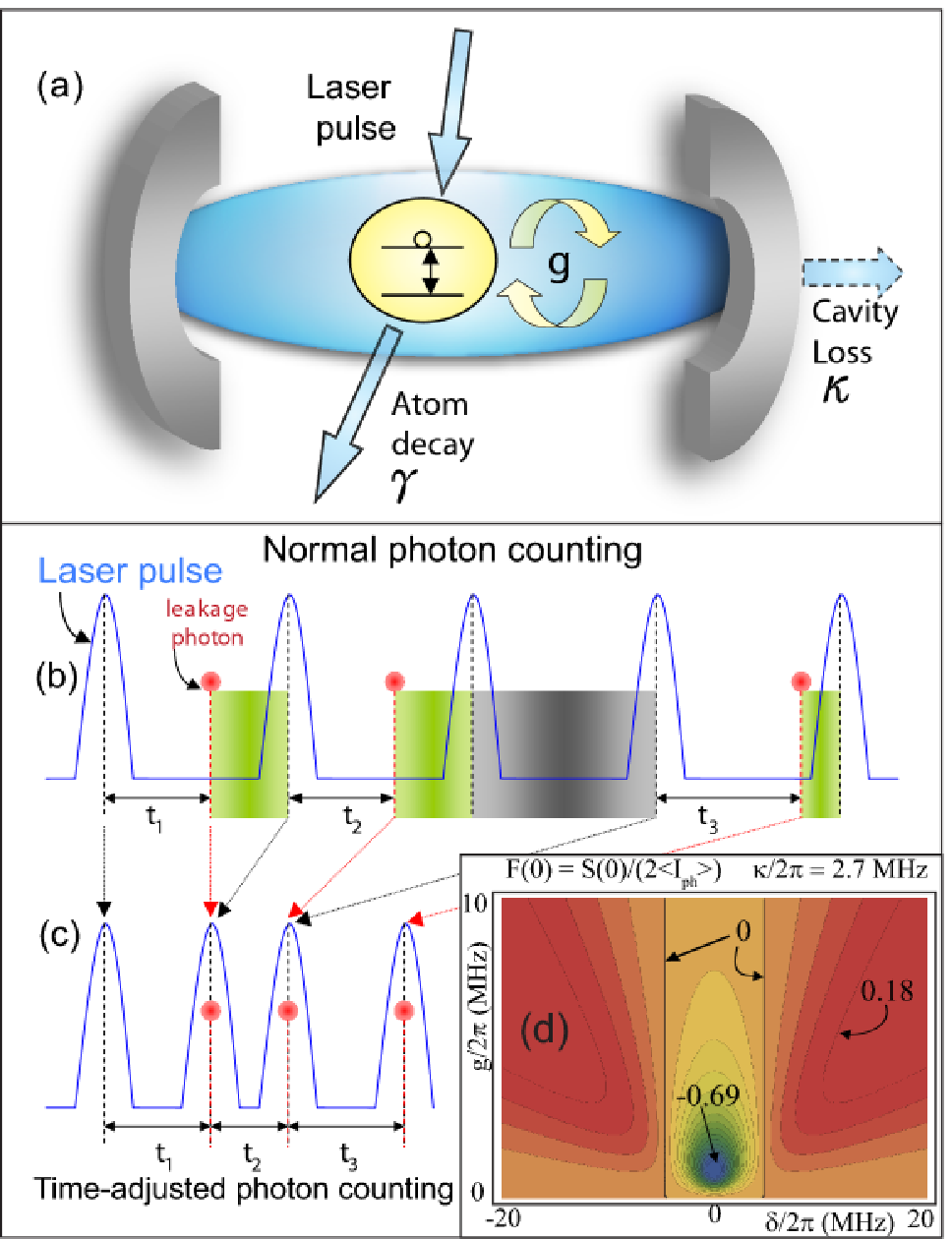}
\caption{{} (a) Schematic diagram of a QED system. Vacuum Rabi
oscillations can occur if the atom-photon coupling strength $g$
overwhelms the loss rate $\protect\kappa $ of cavity photons and
the emission rate $ \protect\gamma $ into other modes.
 (b) Normal photon counting: Periodic ultra-short laser pulses
excite the atom faster than all other time scales. The
single-photon detector records the arrival time $t_{n}$ of a
photon decaying out of the cavity with respect to each pulse. The
grey area in (b) means that no photon is detected due to detector
inefficiency. (c) Time-adjusted photon counting: The time (shown
in green) between a photon detection event (in red) and the
subsequent laser pulse is eliminated, moving the time of each
laser pulse to the time of the previous photon count. Any periods
(in grey) of no-photon detection are eliminated. In this manner
the system has an effective instantaneous feedback: once a photon
is detected, a laser pulse immediately drives the atom to the
excited state. (d) The zero-frequency component of the power
spectrum $F(0)$ is positive in the presence of coherent VRS
(vacuum Rabi splitting), \emph{even if the coupling $g$ is very
small}. In (d), brown is maximum, blue is minimum, and the black
vertical lines show $F(0)=0$.}
\end{figure}

\begin{center}
\begin{table*}[th]
{\small \hfill {}
\begin{tabular}{|c|c|c|}
\hline
\textbf{System} & \textbf{Double Quantum Dot} & \textbf{Cavity QED} \\ \hline
\textbf{Carrier} & Electrons & Photons \\ \hline
\textbf{Ground State} & $|R\rangle =$ electron in the right dot & $%
|g,1\rangle =|$ ground state atom, $1$ photon $\rangle $ \\ \hline
\textbf{Excited State} & $|L\rangle =$ electron in the left dot & $%
|e,0\rangle =|$ excited atom, $0$ photons $\rangle $ \\ \hline
\textbf{Energy difference} $\triangle $$E$ & $E_{L}-E_{R}$ & $\delta
/2\;=\;(\omega -\mu )/2$ \\ \hline
\textbf{Rabi Rate} & Tunnelling amplitude $T$ & Atom-Photon coupling $g$ \\
\hline
\textbf{Input Rate} & Tunnelling rate $\Gamma _{L}\rightarrow \infty $ &
Laser pulses with time-adjusted shift \\ \hline
\textbf{Output Rate} & Tunnelling rate $\Gamma _{R}$ & Cavity loss rate $%
\kappa $ \\ \hline
\textbf{Quantum noise signature} & Super-Poissonian \ $F_{e}(\omega
\rightarrow 0)>1$ & Non-negative \ $F_{\mathrm{ph}}(\omega \rightarrow 0)>0$
\\ \hline
\textbf{Extended LG inequality} & $|2\langle I(t+\tau )I(t)\rangle -\langle
I(t+2\tau )I(t)\rangle |\leq \Gamma _{R}\langle I(t)\rangle $ & $%
|2g^{(2)}(t,t+\tau )-g^{(2)}(t,t+2\tau )|\leq \langle a^{\dagger
}(t)a(t)\rangle ^{-1}$ \\ \hline
\end{tabular}%
} \hfill {}
\caption{Comparison between the properties of the cavity QED system studied
here and a double quantum dot.}
\end{table*}
\end{center}

\section{Results}

\subsection{Standard photon-counting}

A way to %measure the decay
%time of an atom in a cavity, or to
produce$^{1}$ single photons from a cavity is the following:
Ultra-short laser pulses with a given time constant are applied to
the atom. The single-photon detector records the arrival time
$t_{n}$ of a photon decaying out of the cavity with respect to
each pulse, as shown Fig.~1. A normalized histogram of detection
times reveals$^{1}$ photon anti-bunching
and oscillations due to the atom-cavity coupling. %A
%histogram of these photon counts gives the decay curve of the
%system.

Neglecting the emission rate $\gamma $ into other modes, the
normal cavity
QED system$^{4}$ can be described by the Markovian master equation (setting $%
\hbar =1$ throughout),
\begin{equation}
\dot{\rho}=\mathcal{W}_{c}[\rho ]=-i[H_{c},\rho ]+L_{c}[\rho ],  \label{ME}
\end{equation}%
where
\begin{equation}
L_{c}[\rho ]=\kappa a\rho a^{\dagger }-\frac{\kappa }{2}[a^{\dagger }a\rho
+\rho a^{\dagger }a],  \label{LC}
\end{equation}%
and
\begin{equation}
H_{c}=\nu a^{\dagger }a+\frac{\omega }{2}\sigma _{z}+g(\sigma _{-}a^{\dagger
}+\sigma _{+}a).  \label{HC}
\end{equation}%
Here $\omega $ is the atomic level splitting, $\nu $ is the cavity
frequency, and $\kappa $ the cavity loss rate via which we acquire
photons. Here, $a^{\dagger }$ $(a)$ denotes the creation
(annihilation) operator of a cavity photon. The atomic operators
are defined as: $\sigma _{z}\equiv |e\rangle \langle e|-|g\rangle
\langle g|$, $\sigma _{-}\equiv |g\rangle \langle e|$, and $\sigma
_{+}\equiv |e\rangle \langle g|$, where $|e\rangle $ and
$|g\rangle $ denote the excited and ground states, respectively.
The atomic polarization decay $\gamma $ can be easily included in
this analysis, but for simplicity we neglect it here. Furthermore,
we omit variations in the coupling strength $g$ that can occur
between each pulse. This could be an important factor, which can
be dealt with by numerically integrating our final result over a
Gaussian spread in $g$, or by including an additional dephasing
term in the master equation.

\subsection{Time-adjusted photon counting}

Now, rather than ``collating'' data in the above manner, as shown
in Fig.~1(b), we propose to perform a time-adjusted analysis of
the photon data, as shown in Fig.~1(c). Namely, the time between a
photon-detection event and the next laser pulse (shown in green in
the figure) is eliminated, moving the time of each laser pulse to
the time of the previous photon count; and any periods of time
with no-photon detection are eliminated from the data set. Thus,
the system can then be viewed as one with instantaneous feedback
that maintains one excitation in the combined atom-cavity basis.

We show this rigorously by using the feedback formalism in
refs.~17-18 (refer to Methods for the detailed proof). Ultimately,
this transforms the original equation of motion [equations~(1-3)]
into an equation of motion in the pseudo-spin two-state basis
[defined as $\tilde{\sigma}_{z}=|e,0\rangle \langle
e,0|-|g,1\rangle \langle g,1|$, where $e$ $(g)$ represents the
atom in its excited (ground) state, and $1$ $(0)$ represents a
single (no) photon in the cavity]. The restricted master equation
(where feedback has been implicitly introduced, see Methods), is
\begin{eqnarray}
\dot{\rho} &=&-i\bigg[\frac{\nu }{2}+\frac{\delta }{2}\tilde{\sigma}_{z}+g%
\tilde{\sigma}_{x},\rho \bigg]  \notag \\
&+&\kappa \tilde{\sigma _{+}}\rho \tilde{\sigma _{-}}-\frac{\kappa }{2}\bigg[%
\tilde{\sigma _{-}}\tilde{\sigma _{+}}\rho +\rho \tilde{\sigma _{-}}\tilde{%
\sigma _{+}}\bigg],
\end{eqnarray}%
where $\delta =\omega -\nu $ is the detuning between the atom and
the cavity. This restricted-basis equation of motion is equivalent
to a two-level atom undergoing resonance fluorescence in free
space$^{19}$. Here, the two-level
atom is represented by the combined atom-cavity states $|g,1\rangle $ and $%
|e,0\rangle $. The coherent input field is the natural atom-cavity
interaction, and the time-adjusted photon counting gives an
effective decay from $|g,1\rangle $ to $|e,0\rangle $, by
eliminating the no-excitation state $|g,0\rangle $. The
time-adjusted data-set then represents a single trajectory in the
ensemble described by this new equation of motion.

\subsection{Second-order correlation function}

From this simple two-state model, the ensemble-averaged measurements of the
photon output from the cavity can be easily calculated. In particular, the
second-order correlation function,
\begin{equation}
g^{(2)}(t,t+\tau)=\frac{\langle a^{\dagger }(t)a^{\dagger }(t+\tau )a(t+\tau
)a(t)\rangle }{\langle a^{\dagger }(t)a(t)\rangle ^{2}}
\end{equation}
is found to be
\begin{equation}
g^{(2)}=\frac{e^{-\alpha t}}{8\Theta }\left[ 3\kappa -4\Theta +8\Theta
e^{\alpha t}-4\alpha e^{2\Theta t}\right] ,
\end{equation}
where $\alpha= \frac{3\kappa}{4}+\Theta$, $\Theta =\sqrt{\kappa
^{2}/16-4g^{2}}$, and $\delta=0$ for convenience.
%which is identical to that for single atom undergoing resonant fluorescence.
%As in the feedback experiment of Smith \textit{et al}$%
%^{11}$, symmetry of this function is lost. Moreover
It is important to point out that one cannot define a correct first-order
correlation function $G^{(1)}(t,t+\tau) = \langle a^{\dagger}(t)a(t+\tau)
\rangle$ with this two-state model, unless one performs a full numerical
simulation using a trace-preserving feedback operator (see Methods), or one
retains the incoherent transition through the $| g,0 \rangle$ state in the
equation of motion.

\subsection{Analogy with electron transport}

As discussed earlier, our goal is to show that this simple model is
equivalent to the electron transport through a solid-state double quantum
dot, and then take advantage of common tools from transport theory. In the
transport regime, a double quantum dot is connected to electronic reservoirs
with tunnelling rates $\Gamma _{L}$ and $\Gamma _{R}$. If one assumes the
strong Coulomb-blockade regime, i.e., the charging energy is much larger
than other parameters, then one only needs to consider a single level in
each dot. One can then define the three-state basis: $\left| L\right\rangle $%
, $\left| R\right\rangle $ and, $\left| 0\right\rangle $
representing an electron in the left-dot, the right-dot, and the
empty state, respectively. The DQD
Hamiltonian is written as%
\begin{equation}
H_{d}=E_{L}\left| L\right\rangle \left\langle L\right| +E_{R}\left|
R\right\rangle \left\langle R\right| +T\left| L\right\rangle \left\langle
R\right| +T\left| R\right\rangle \left\langle L\right| ,
\end{equation}
where $E_{L}$ $(E_{R})$ is the energy for the left (right)-dot
level and $T$ is the coherent tunnelling amplitude between them.

The density matrix $\rho (t)$ of the DQD satisfies
\begin{equation}
\frac{d}{dt}\rho (t)=\mathcal{W}_{d}[\rho (t)]=-i[H_{d},\rho (t)]+L_{d}[\rho
(t)].
\end{equation}%
The $L_{d}$ term in equation (8) contains the transport properties
and dissipation within the device,
\begin{eqnarray}
L_{d}[\rho (t)] &=&-\frac{\Gamma _{L}}{2}[s_{L}s_{L}^{\dagger }\rho
(t)-2s_{L}^{\dagger }\rho (t)s_{L}+\rho (t)s_{L}s_{L}^{\dagger }]  \notag \\
&-&\frac{\Gamma _{R}}{2}[s_{R}^{\dagger }s_{R}\rho (t)-2s_{R}\rho
(t)s_{R}^{\dagger }+\rho (t)s_{R}^{\dagger }s_{R}],
\end{eqnarray}%
where $s_{L}=\left| 0\right\rangle \left\langle L\right| ,$
$s_{L}^{\dagger }=\left| L\right\rangle \left\langle 0\right| ,$
$s_{R}=\left| 0\right\rangle \left\langle R\right| ,$ and
$s_{R}^{\dagger }=\left| R\right\rangle \left\langle 0\right| $.
One can calculate the current of electrons leaving the device
using a current super-operator (e.g., for the
junction on the right, and setting the electric charge $e=1$ throughout) $%
\widehat{I}_{R}\rho (t)=\Gamma _{R}\left| 0\right\rangle
\left\langle R\right| \rho (t)\left| R\right\rangle \left\langle
0\right| $. The steady-state current of the right junction is
then$^{20,21}$
\begin{eqnarray}
\left\langle I_{s}\right\rangle
&=&\mathrm{Tr}\left(\widehat{I}_{R}\rho _{0}\right)
\notag \\
&=&\frac{\Gamma _{R}T^{2}}{\Gamma _{R}^{2}/4+\varepsilon ^{2}+T^{2}(2+\Gamma
_{R}/\Gamma _{L})},
\end{eqnarray}%
where $\varepsilon =E_{L}-E_{R}$ is the energy difference between
the two dots. The shot noise $S_{e}(\omega )$ of this
device$^{22,23}$ can also be easily obtained,
\begin{eqnarray}
S_{e}(\omega ) &=&\int_{-\infty }^{\infty }d\tau e^{i\omega \tau }\langle
\lbrack \delta I_{R}(t),\delta I_{R}(t+\tau )]_{+}\rangle _{t\rightarrow
\infty }  \notag \\
&+&2\langle I_{s}\rangle \delta (\tau ),
\end{eqnarray}%
where the fluctuating right-junction current is $\delta I_{R}(t)=\widehat{I}%
_{R}(t)-\langle I_{s}\rangle $, and the self-correlation term
$2\langle I_{s}\rangle \delta (\tau )$ represents the correlation
of a tunnelling event with itself$^{20}$. The shot noise
(zero-frequency noise) is found to be$^{22,23}$,
\begin{eqnarray}
F_{e} &=&S_{e}(0)/2e\left\langle I_{s}\right\rangle =\bigg\{1- \\
&&8T^{2}\Gamma _{L}\frac{4\varepsilon ^{2}(\Gamma _{R}-\Gamma _{L})+3\Gamma
_{L}\Gamma _{R}^{2}+\Gamma _{R}^{3}+8\Gamma _{R}T^{2}}{[\Gamma _{L}\Gamma
_{R}^{2}+4\Gamma _{L}\varepsilon ^{2}+4T^{2}(\Gamma _{R}+2\Gamma _{L})]^{2}}%
\bigg\}.  \notag
\end{eqnarray}

To reduce the double quantum dot problem, from a three-state basis
to a two-state one, we take the limit of $\Gamma _{L}\gg \Gamma
_{R},T,\varepsilon $. This allows us to eliminate the $|0\rangle $
empty state, so that as an electron tunnels out of the right
junction, an electron immediately tunnels through the left one;
mimicking the effective two-level behavior of the cavity QED
system. Interestingly, this limit gives exact results for all the
stationary currents, but only returns the correct time-dependence
for the right-junction current, in analogy with the
inability for the restricted basis cavity-QED model to correctly construct $%
G^{(1)}$. In this case, the super-operator for the right-junction
current becomes, $\widehat{I}_{R}\rho (t)=\Gamma _{R}\left|
L\right\rangle \left\langle R\right| \rho (t)\left| R\right\rangle
\left\langle L\right| $. Similarly, the zero state is eliminated
from the $L_{d}$ term so that only one tunnelling rate,
$\Gamma_R$, remains. In this limited basis, equations (8) and (9)
are equivalent to equation (4). Table I lists how the various
parameters correspond to one another in the two different systems.

Of particular interest to us is how the right-junction
second-order current-correlation function in the large $\Gamma_L$
limit is equivalent to the second-order photon correlation
function $g^{(2)}(t,t+\tau )$ discussed above. This is because (in
the limit $\Gamma _{L}\gg \Gamma _{R},T,\varepsilon $), \beq
\langle
I_{R}(t+\tau )I_{R}(t)\rangle =\mathrm{Tr}[\hat{I}_{R}e^{\mathcal{W_d}\tau }%
\hat{I}_{R}\rho (t)]\nonumber \\ =\Gamma _{R}^{2}\mathrm{Tr}[|L\rangle \langle R|\{e^{%
\mathcal{W_d}\tau }|L\rangle \langle R|\rho (t)|R\rangle \langle
L|\}|R\rangle \langle L|],\eeq where each super-operator acts on
those to the right. The corresponding correlation function for
$g^{(2)}(t,t+\tau )$ for the cavity QED system, in the reduced
basis we discussed earlier, is defined as \beq%
g^{(2)}(t,t+\tau )\langle a^{\dagger }(t)a(t)\rangle
^{2}&=&\langle a^{\dagger }(t)a^{\dagger }(t+\tau )a(t+\tau
)a(t)\rangle
\nonumber\\&=&\mathrm{Tr}[\tilde{\sigma}_{-}\tilde{\sigma}_{+}e^{\mathcal{W_c}\tau }\tilde{\sigma}%
_{+}\rho (t)\tilde{\sigma}_{-}],\eeq where we have adopted the
traditional input-output formalism to define the photon intensity
in terms of the internal pseudo-spin operators $\tilde{\sigma}$.
It is easy to see that the current-correlation measurement can be
made equivalent to the second-order photon-intensity measurement
simply by multiplying by a factor of $\kappa
^{2}$. %$\ex{a^{\dagger}(t)a^{\dagger}(t+\tau)a(t+\tau)a(t)}$

In summary, the ensemble-averaged photon statistics from a
periodically-pulsed cavity\ QED system, following appropriate
time-adjustments, has the same properties as the transport of electrons
through a double quantum dot. As an example of the power of this apparently
simple analogy, summarized in table I, we examine two tests for quantum
behavior in double quantum dots (non-negative shot noise, and a special case
of the Leggett-Garg inequality), and show how these two tests can be applied
to the time-adjusted cavity QED system.

\subsection{Tests of quantum-ness}

\subsubsection{Super-Poissonian shot noise}

It has been argued (and shown experimentally) that
super-Poissonian shot noise can be observed in double quantum dots
only if there is coherent quantum tunnelling between the two
dots$^{24,25}$. If $\Gamma _{L}$ is larger than $\Gamma _{R}$, the
second term in equation (12) becomes negative and produces a
super-Poissonian value: $S(0)/2e\left\langle I_{s}\right\rangle
>1$. In our analogy, $\Gamma _{L}$ is always much larger than
$\Gamma _{R}$ (it is effectively infinite). Using the
correspondence between electron current and photon intensity, we
can define an effective fluctuating photon-intensity
noise-spectrum
\begin{equation*}
S_{\mathrm{ph}}(\omega )=2\mathrm{Re}\Bigg[\int_{0}^{\infty }\!\!\!\!d\tau
\;e^{i\omega \tau }\langle I_{\mathrm{ph}}\rangle ^{2}\bigg(g^{(2)}(t,t+\tau
)-1\bigg)\Bigg],
\end{equation*}%
where the effective photon current is $\langle I_{\mathrm{ph}}\rangle
=\kappa \langle a^{\dagger }(t)a(t)\rangle $, and the subscript ``$\mathrm{ph%
}$'' represents the photon analogy to typical electron transport
measurements. In the photon case there is no self-correlation term. We can
easily calculate a photon current Fano factor using the same technique used
for DQDs, and find%
\begin{equation}
F_{\mathrm{ph}}=\frac{S_{\mathrm{ph}}(0)}{2\left\langle I_{\mathrm{ph}%
}\right\rangle }=-\frac{8g^{2}(3\kappa ^{2}-\delta ^{2})}{(8g^{2}+\kappa
^{2}+\delta ^{2})^{2}}.
\end{equation}%
From this equation one can easily see that super-Poissonian noise in the
transport case corresponds to positive noise in the photon case ($F_{\mathrm{%
ph}}>0$). This can occur if $3\kappa ^{2}<\delta ^{2}$, otherwise
the `shot noise' for photons is negative. As mentioned above, in
the DQD electron transport case, the super-Poissonian noise is
only obtained for coherent coupling between the two dots. For
classical sequential tunnelling between two dots, the result turns
out to be solely sub-Poissonian$^{24,25}$. For the time-adjusted
single-photon cavity-QED system we consider here, coherent Rabi
oscillations between the atom and cavity photon states produce a
positive zero-frequency component in the shot-noise spectrum. This
is clearly indicated in Fig.~1(d), using parameters akin to those
in the experiment in ref.~1. Even for $g\leq \kappa $, the
spectrum remains positive. Only for $\kappa \gg g$ does the whole
range of $F_{\mathrm{ph}}(0)$, as a function of $\delta $, become
negative.

In quantum optics, it is well known that a single-photon source
has sub-Poissonian behavior in the second-order correlation
function as the time $\tau$ between measurements goes to zero:
$g^{(2)}(t,t+\tau )_{\tau \rightarrow 0}\rightarrow 0$. This is
termed anti-bunching$^{26,27}$. One can then call the light
observed in this way as `non-classical': it cannot be obtained
from a thermal or coherent state source. Equation (13) implies a
secondary
criterium for the `quantum-ness' of the observed light from such a \emph{%
single-photon source}: if the correlation function is conditioned by quantum
coherent oscillations (vacuum Rabi splitting), one should see a positive
value for the \emph{zero-frequency} limit of the photon noise spectrum.

\begin{figure}[h]
\includegraphics[width=\columnwidth]{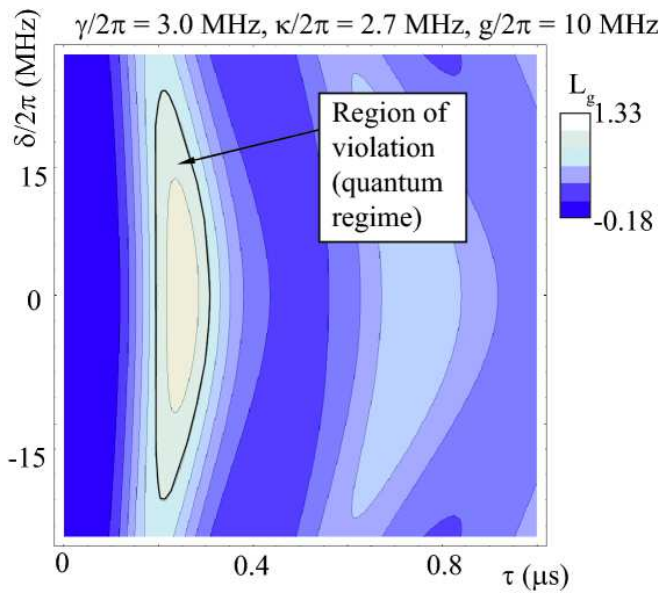}%fig4new.eps}
\caption{{} A violation of the extended Leggett-Garg inequality
(15) for typical parameters in single-photon cavity QED
experiments$^{2,3}$, using time-adjusted photon statistics. The
parameters used here are: $\protect\kappa/2\protect\pi=$2.7 MHz,
$g/2\protect\pi=$10 MHz, and the variation of the detuning
$\protect\delta/2\protect\pi $ can be up to 20 MHz. The violation
of the inequality is indicated by the grey ``island'' inside the
black contour line. }
\end{figure}
\begin{figure}[h]
\includegraphics[width=\columnwidth]{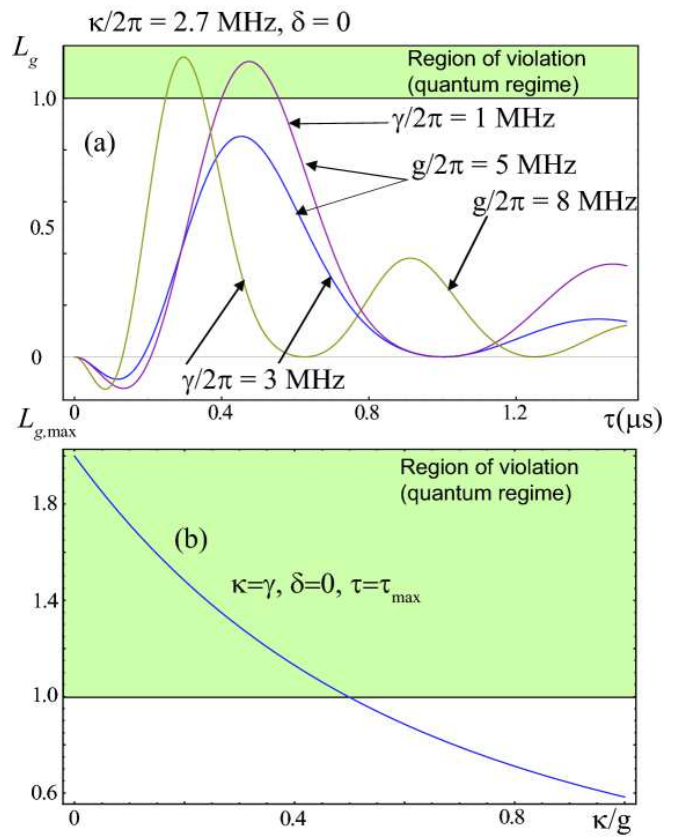}%fig5.eps}
\caption{{} (a) The non-adjusted photon-counting statistics also
exhibit a violation of the Leggett-Garg inequality (15). The
combined
effect of the atom polarization decay and the cavity decay rates ($\protect%
\gamma/ 2\protect\pi=3$ MHz and $\protect\kappa/2\protect\pi= 2.7$ MHz),
prevent a violation being seen in the data in ref.~1. However, a slight
decrease in the polarization decay rate $\protect\gamma$; or an increase in
the coupling strength $g$, combined with the ultra-short laser pulses in
ref.~1, should reveal a violation of the inequality (15). (b) Using the same
model (without time-adjusted counting), one can estimate the ratio between
the coupling strength $g$ and the dissipation ($\protect\kappa$, $\protect%
\gamma$) needed to observe a violation of (15). This is approximately given
by the vacuum Rabi splitting VRS$\,= 4g/(\protect\kappa+\protect\gamma) > 4 $%
. In (b) we have chosen $\protect\kappa=\protect\gamma$.}%
%
%The
%condition for violation in for time-adjusted statistics is almost
%identical.}
\end{figure}

\subsubsection{Extension of the Leggett-Garg Inequality}

To further clarify the quantum signatures in these correlation
functions, we turn to an extension$^{15}$ of the Leggett-Garg
inequality$^{16}$. Recently, we derived our extended inequality
based on measurements of the fluctuating current through a DQD
device$^{15}$. Since the current is essentially an invasive
measurement, we showed that the inequality only applies for
systems with a state space of three or less states, and under the
assumption of irreversible transport into the right reservoir (see
Methods). For the cavity QED system, these two assumptions become
that of being in the single-excitation manifold, and the
irreversible loss of photons once they leave the cavity,
respectively. Thus, the current inequality of ref.~15,
\begin{equation}
|L_{I}(t)|\equiv |2\langle I(t+\tau )I(t)\rangle -\langle I(t+2\tau
)I(t)\rangle |\leq \Gamma _{R}\langle I(t)\rangle ,
\end{equation}%
becomes a second-order photon-correlation inequality,
\begin{equation}
|L_{g}(t)|\equiv |2g^{(2)}(t,t+\tau )-g^{(2)}(t,t+2\tau )|\leq {\langle
a^{\dagger }(t)a(t)\rangle }^{-1}.
\end{equation}%
Superficially, this inequality is equivalent to the Leggett-Garg
inequality in the stationary limit$^{16}$. However, as mentioned
above, the observation of a photon is an invasive measurement in
terms of the cavity-atom state, and thus, strictly speaking, this
inequality no longer can be discussed in terms of distinguishing
theories obeying macroscopic realism from quantum mechanics: the
original goal of the work of Leggett and Garg. Here this
inequality only distinguishes quantum dynamics from those given by
a classical rate equation. See the methods section, and ref.~15,
for more details. Like the non-negative shot-noise feature, this
inequality reveals a more nuanced way of understanding whether
certain photon statistics have quantum characteristics, beyond
those indicated by anti-bunching alone.

In Fig.~2, we show how the violation of this inequality occurs for a typical
cavity QED experiment, using$^{1-4}$ $\kappa /2\pi =$2.7 MHz, $\gamma /2\pi
=3$ MHz, $g=$10 MHz. As seen in Fig.~2, the violations of the inequality are
easily observable and appear in a wide range of detuning $\delta $.

\subsubsection{Extended inequality with standard photon statistics}

One can also apply the extended$^{15}$ Leggett-Garg inequality to
the photon statistics without time-adjustment. In this case, a
histogram of the photon counts as a function of time (after the
initial excitation of the atom) is equivalent to the second-order
correlation function of the atom-cavity system with one excitation
and cavity decay, but no further time-dependent excitations. See
refs.~1-4 for clear examples of such statistics.

Using a simple model of the Bochmann et al's experiment$^{1}$ (using
equations (1-3), and the initial state $\rho _{0}=|g,1\rangle \langle g,1|$%
), we have found that their experiment does not violate the
inequality (15) (the bound is now set by the choice of initial
state, see methods). A factor of two decrease in the atomic
polarization decay rate, or a correspondingly stronger coupling
strength, should reveal a violation. We illustrate this in
Fig.~3a, again using their parameters; though we omit dephasing
because of variations in coupling $g$. It is easy to see that a
violation of (15) should be possible with minor improvements in
system parameters. To give a more general bound for parameters
which can cause a violation, in Fig.~3b we show the magnitude of
the violation versus the cavity and atomic losses. This gives a
bound on the Vacuum Rabi Splitting parameters needed to observe a
violation of $2g/[(\gamma +\kappa )/2]>4$. Many realizations of
cavity QED have parameters which exceed this (see, e.g., ref.~5).
We therefore think that this inequality (15) is a useful addition
to the toolbox one can use to test for quantum behavior in optical
systems (see refs.~26,27 for reviews of other common tests).

\section{Discussion}

We have shown how a simple adjustment of the output photon-detection
statistics of a periodically-excited cavity QED system can be described by a
non-equilibrium model, with an exact analogy to electron transport
properties through a double quantum dot. This represents a unification of
the fields of photon-counting statistics and electron-transport statistics.
We then adapted several recent results from transport theory to describe or
test the quantum nature of the photon-statistics being emitted from the
cavity.

We emphasize that not only the current noise, but also the
higher-order moments$^{28,29}$ can be examined with this
time-adjusted scenario.
Moreover, we point out that the same features could be observed in a \emph{%
circuit} QED system, where the artificial atom (qubit) is periodically
excited by some external means, and photons detected with a microwave photon
counter$^{30}$. % Such a
%device has recently been proposed$^{20}$.% and one can expect this will be
%realized in the near future.

\subsection{\textbf{ACKNOWLEDGMENTS}}

NL and YN-C contributed equally to this work. NL is supported by the RIKEN
FPR program. YN-C is supported partially by the National Science Council of
Taiwan under the grant number 98-2112-M-006-002-MY3. FN acknowledges partial
support from the Laboratory of Physical Sciences, National Security Agency,
Army Research Office, National Science Foundation grant No. 0726909,
JSPS-RFBR contract No. 09-02-92114, Grant-in-Aid for Scientific Research
(S), MEXT Kakenhi on Quantum Cybernetics, and Funding Program for Innovative
R\&D on S\&T (FIRST).

\newpage

\section{Methods}

\noindent \textbf{Effective feedback formalism.} First, we describe the
measurement of a single photon (which has leaked from the cavity and is
incident on a photodetector) as $\mathcal{K}(dt)_{1}=\sqrt{dt}\,a$. %,
%i.e. as the annihilation of a photon at a photodetector.
The complementary operator to this one, which is applied when no
photon is detected during the duration $dt$, is
$\mathcal{K}(dt)_{2}=\sqrt{dt}\,a-(iH_{c}+\frac{1}{2}a^{\dagger
}a)dt$. The feedback evolution which is then
applied to the system, given the observation of a photon, is described by $%
\mathcal{O}[...]=|e\rangle \langle g|...|g\rangle \langle e|$;
i.e. the atom is \emph{instantaneously} and \emph{incoherently}
projected into its excited state. In general, this is not a
trace-preserving evolution, as it is not a Liouvillian evolution,
and is thus different from the class of feedback mechanisms
described in refs.~17,18. Alternatively, one could assume a
Liouvillian evolution according to the coherent dynamics of a
laser pulse causing $\pi $-pulse/Rabi oscillations of the atom
from its ground to excited state, via an operator
$e^{-i\mathcal{Z}}$, $\mathcal{Z}=[\sigma _{x},...]$. However, in
the limit when this transition is faster than all other dynamics,
and we are in the single-excitation manifold, then it is
equivalent to $\mathcal{O}[...]$.

The unnormalized density matrix following the detection of a photon at time $%
t$, and evolution due to feedback, is
\begin{equation*}
\tilde{\rho}_{1}(t+dt)=\kappa \mathcal{O}a\rho (t)a^{\dagger }dt.
\end{equation*}%
Our time-adjustment scheme implies that the time delay is far smaller
(effectively zero) than the cavity decay time, and one can assume the fully
Markovian master equation,
\begin{equation*}
\dot{\rho}=-i[H_{c},\rho ]+\kappa \mathcal{O}a\rho a^{\dagger }-\frac{\kappa
}{2}\bigg[a^{\dagger }a\rho +\rho a^{\dagger }a\bigg].
\end{equation*}

If the feedback is via the operator $\mathcal{O}[...]$, this is not a
trace-preserving equation of motion. However, if we restrict ourselves to
the single (lowest) excitation manifold, only the $\left| e,0\right\rangle $%
, $\left| g,1\right\rangle $, and $|g,0\rangle $ states are
important. Then the feedback term becomes $\mathcal{O}a\rho
a^{\dagger }=|e,0\rangle \langle g,1|\rho |g,1\rangle \langle
e,0|$, and this truncated basis is trace-preserving, as the state
$|e,1\rangle $ is decoupled from the equation
of motion, giving $\mathrm{Tr}[\mathcal{O}a\rho a^{\dagger }]=\mathrm{Tr}%
[a\rho a^{\dagger }]=\langle g,1|\rho |g,1\rangle $.

Now the action of the instantaneous feedback is obvious, such that the state
$|g,0\rangle $ in the photon-decay terms is decoupled from the
single-excitation manifold. We can now write our equation of motion purely
in the pseudo-spin two-state basis, defining $\tilde{\sigma}_{z}=|e,0\rangle
\langle e,0|-|g,1\rangle \langle g,1|$,
\begin{eqnarray}
\dot{\rho} &=&-i[\nu +\delta \tilde{\sigma}_{z}+g\tilde{\sigma}_{x},\rho ]
\notag \\
&+&\kappa \tilde{\sigma _{+}}\rho \tilde{\sigma _{-}}-\frac{\kappa }{2}\bigg[%
\tilde{\sigma _{-}}\tilde{\sigma _{+}}\rho +\rho \tilde{\sigma _{-}}\tilde{%
\sigma _{+}}\bigg],  \notag
\end{eqnarray}%
where $\delta =\omega -\nu $ is the detuning between the atom and
the cavity.

To achieve the above, we note the following points. First, the
delay between measurement and feedback is instantaneous. In our
case the delay is zero, as dictated by our adjustment of time
intervals in the data set. Second, the feedback action (i.e.,
laser
pulse) \emph{incoherently} and instantaneously projects the system into the $%
|e,0\rangle $ state. This has recently been achieved by Bochmann et al.$^{1}$
using ultra-short laser pulses. Third, the photon detection is here assumed
to be $100\%$ efficient. This can be effectively achieved by simply omitting
the time intervals where no photons are detected.

\noindent \textbf{Derivation of the extended Leggett-Garg
inequality.} In ref.~15, we extended the Leggett-Garg inequality
to work under the conditions of invasive measurement, but with
additional restrictions. We summarize and reformulate the proof of
that inequality here, but now using the language of cavity-QED.

For the cavity-QED case, we posit that any photon-intensity measurements not
conditioned by quantum dynamics obey
\begin{equation}
|L_{g}(t,t+\tau )|\equiv |2g^{(2)}(t,t+\tau )-g^{(2)}(t,t+2\tau )|\leq {%
\langle a^{\dagger }(t)a(t)\rangle }^{-1}.
\end{equation}%
In the language of an effective photon current, we can write this as
\begin{eqnarray}
|L_{I}(t,t+\tau )| &\equiv &  \label{LI} \\
|2\langle I_{\mathrm{ph}}(t+\tau )I_{\mathrm{ph}}(t)\rangle &-&\langle I_{%
\mathrm{ph}}(t+2\tau )I_{\mathrm{ph}}(t)\rangle |\leq \kappa \langle I_{%
\mathrm{ph}}\rangle ,  \notag
\end{eqnarray}%
where $\kappa $ is the rate of photons leaking from the cavity, $I_{\mathrm{%
ph}}(t)\equiv I_{\mathrm{ph}}(t=0)$ and $\langle I_{\mathrm{ph}}(t)\rangle $
is the average photon current of the initial state. Hereafter we omit the $t$
variable. In the master equation approach, the current operator translates
into a ``jump'' super-operator, and equation~(\ref{LI}) thus represents an
inequality concerning transitions in the system, and not static properties.
Thus it is obviously suitable for application to single-photon measurements,
which give us information about a change in the state of the cavity-QED
system. As described in the text, the photon current super-operator acts as
before $\hat{I}_{\mathrm{ph}}[\rho ]=\kappa \tilde{\sigma}^{+}\rho \tilde{%
\sigma}^{-}$, such that the average current is $\langle I_{\mathrm{ph}%
}\rangle =\mathrm{Tr}\left\{ \hat{I}_{\mathrm{ph}}\rho \right\} $ and the
correlation function of interest is obtained as $\langle I_{\mathrm{ph}%
}(\tau )I_{\mathrm{ph}}\rangle =\mathrm{Tr}\left\{ \hat{I}_{\mathrm{ph}}e^{%
\mathcal{L}\tau }\hat{I}_{\mathrm{ph}}\rho _{0}\right\} $. For our
time-adjusted case, the stationary distribution is chosen as the
initial state. For the non-time-adjusted photon statistics, one
chooses $\rho _{0}=|g,1\rangle \langle 1,g|$.

In these terms, the inequality expression can be written as $L_{I}(\tau )=%
\mathrm{Tr}\left\{ \hat{I}_{\mathrm{ph}}\left( 2e^{\mathcal{L}\tau }-e^{2%
\mathcal{L}\tau }\right) \hat{I}_{\mathrm{ph}}\rho _{0}\right\} .$ If the
cavity-QED system contains no coherent quantum dynamics, $\hat{I}_{\mathrm{ph%
}}$ is the $3\times 3$ matrix with elements $\hat{I_{\mathrm{ph}}}_{\alpha
\beta }=\kappa \delta _{\alpha ,0}\delta _{\beta ,\mathcal{P}}$, where the
indices $0=|g,0\rangle ,\mathcal{P}=|g,1\rangle ,\mathcal{A}=|e,0\rangle $.
Thus using the Chapman-Kolgomorov equation, we have
\begin{equation*}
L_{I}(\tau )=\kappa ^{2}P_{\mathcal{P}}(0)\bigg[\Omega _{\mathcal{P}0}\left(
2-\Omega _{00}-\Omega _{\mathcal{P}\mathcal{P}}\right) -\Omega _{\mathcal{P}%
\mathcal{A}}\Omega _{\mathcal{A}0}\bigg].
\end{equation*}%
Where $\Omega $ represents the matrix elements of the propagator. For a
general Markov stochastic matrix, $\Omega $, the maximum of $L_{I}$ is
\begin{equation*}
\mathrm{max}\{L_{I}\}=2\kappa ^{2}P_{\mathcal{P}}(0).
\end{equation*}%
However, the rate equation form $\Omega (\tau )=\exp {\mathcal{L}\tau }$
furnishes us with a further constraint. Maximizing $L_{I}(\tau )$ with
respect to time, from $\dot{L}_{I}=0$ and $\dot{\Omega}=\mathcal{L}\Omega $,
we find that the maximum of $L_{I}$ occurs when $\Omega _{00}+\Omega _{%
\mathcal{P}\mathcal{P}}=1$ and $\Omega _{\mathcal{P}0}=1$, giving
\begin{equation*}
\mathrm{max}\{L_{I}\}=\kappa ^{2}P_{\mathcal{P}}(0)=\kappa \langle
I_{ph}\rangle .
\end{equation*}%
This result relies on the form of the jump operator and the absence of
re-absorbtion of photons by the cavity, i.e. $\mathcal{L}_{\mathcal{P}0}=0$.
%\hfill\ensuremath{\Box}
These requirements mean that we must always be in the single-excitation
regime, and that once a photon leaves the cavity and is measured, it cannot
return. Fortunately, this is implicit in the definition of destructive
photon measurement.

\end{document}